\documentclass{aa} 
\usepackage{natbib}
\usepackage{graphicx}
\usepackage{txfonts}
\usepackage{lipsum}
\usepackage{subcaption}       
\usepackage{lscape}                              
\usepackage{placeins}          
\usepackage{hyperref}
\usepackage{xcolor}
\usepackage{multirow}
\hypersetup{colorlinks=true, linkcolor=orange, citecolor=blue, filecolor=cyan, urlcolor=magenta}
\def\rmd{{\mathrm d}}
\newcommand{\vect}[1]{\mathbf{#1}}
\newcommand{\msun}{h^{-1} M_{\odot}}
\newcommand{\kms}{\rm km\,s^{-1}}
\newcommand{\avg}[1]{\langle{#1}\rangle}
\newcommand{\abs}[1]{\left\vert{#1}\right\vert}

\begin{document}

  \title{Approximating the peculiar velocity distribution of dark matter halos with Tsallis statistics}
   
  \author{Jun Pan\inst{1}\corrauth{jpan@nao.cas.cn} \and Ming Li\inst{2}\email{mingli@nao.cas.cn}}

  \institute{Chinese Academy of Sciences South America Center for Astronomy, National Astronomical Observatories, CAS, Beijing, 100101, People's Republic of China
  \and
   Key Laboratory for Computational Astrophysics, National Astronomical Observatories, Chinese Academy of Sciences, Beijing 100101, People's Republic of China}

  \date{Received XXX XX, 2026}

  \abstract
   {Dark matter halos, which host galaxies and galaxy clusters, have peculiar velocities far from thermal equilibrium. Characterizing the nonlinear and non‑Gaussian features of this velocity distribution improves our understanding of the gravitational evolution of the cosmic web and supports related cosmological applications.}
   {We endeavor to establish a connection between the peculiar velocity distribution of halos and nonequilibrium statistical mechanics, with the objective of obtaining a model that is both concise and accurate for practical applications.}
    {We extracted halo samples from large N-body simulations and performed maximum‑likelihood fits to the peculiar velocity distributions using a two‑parameter Tsallis model, derived from non‑extensive statistical mechanics. On the theoretical side, we reformulated the halo distribution in the superstatistics framework by means of a generalized Gram–Charlier expansion based on the gamma distribution.}
   {For halo peculiar velocities below $\sim 1000\ \kms$, the Tsallis model achieves $5\%$ accuracy over $0\lesssim z\lesssim2$, with performance improving toward lower redshifts. Our results show that the halo velocity distribution becomes increasingly non-Gaussian and departs further from equilibrium over time. The best-fit parameters depend only weakly on mass, though low-mass halos exhibit slightly stronger non-Gaussianity. The two parameters, especially the velocity dispersion, offer promising probes of cosmological parameters. Theoretically, we find that in general the halo peculiar velocity distribution function is expressible as a superposition of a series of Tsallis distribution functions, while simulation results demonstrate that the zeroth-order approximation, namely a single Tsallis function, already achieves sufficient accuracy.}
   {The one-point statistics of halo peculiar velocities at low redshifts can be effectively characterized by the Tsallis distribution.}


  \maketitle
  \nolinenumbers

\section{Introduction}

In the current standard paradigm of cosmic large-scale structure, dark matter halos play a central role as hosts of galaxies, groups, and clusters. Their phase-space distribution is therefore of great importance for studies of galaxy formation and evolution, as well as for cosmology \citep[see, e.g., the reviews of][]{CooraySheth2002, WechslerTinker2018, AsgariEtal2023}. \citet{CroftEftathiou1994} pointed out that the peculiar velocities of clusters can be measured more accurately than those of galaxies, and that the correspondence of dark halos to galaxy clusters facilitates the convenient application of linear or quasi-linear theoretical frameworks. Accordingly, beyond the most commonly used statistical descriptors of the spatial clustering of dark halos—such as correlation functions or power spectra—it is now well recognized that statistics of halo peculiar velocities provide valuable insights into the underlying cosmology and gravitational theory \citep{TsagasEtal2026}. For example, the pairwise velocity distribution is directly linked to the detection of the kinematic Sunyaev–Zel'dovich effect \citep{kSZ1980, Birkinshaw1999} through pairwise statistics \citep[e.g.,][]{HandEtal2012, AdeEtal2016, BernardisEtal2017, CalafutEtal2021, LiEtal2024}, to the analysis of redshift-space distortions \citep[e.g.,][]{Kaiser1987, Fisher1995, Scoccimarro2004, Tinker2007, ZhangEtal2013, KuruvillaPorciani2018}, and to numerous other applications \citep[e.g.,][]{HellwingEtal2014, MaEtal2015, DupuyEtal2019, ZhangEtal2025}. 

The work presented here focuses primarily on the one‑point distribution function of halo peculiar velocities. In principle, a complete statistical description of the peculiar velocity field requires the full hierarchy of multipoint distribution functions. Within this hierarchy, the pairwise velocity distribution is a two‑point function, whereas the one‑point distribution function is the lowest‑order and most fundamental. It serves as the foundation for studying multipoint velocity distribution functions and related physical effects, such as redshift distortion \citep{TNS2010, ZhangEtal2013}.

With advances in modern observational techniques, it is now possible to compile large peculiar velocity samples of clusters and to extract the one‑point distribution from cosmic microwave background observations and galaxy spectroscopic surveys \citep[e.g.,][]{AghanimEtal2018, KourkchiEtal2020, WangEtal2021}. Over the years, numerous attempts have been made to use the distribution function or the root‑mean‑square peculiar velocity of clusters to constrain cosmological models \citep[e.g.,][]{CroftEftathiou1994, BO1996, MoscardiniEtal1996, Watkins1997, BorganiEtal1997, PeelKnox2003, MastersEtal2006}. Many of these studies assumed that the peculiar velocities of halos evolved linearly from Gaussian initial conditions. The problem, however, is that dark matter halos are large gravitationally self‑interacting objects and are not analogous to an ideal gas in thermal equilibrium. The peculiar velocity distribution differs significantly from the canonical Boltzmann–Gibbs statistics \citep{SaslawEtal1990}. Simple extrapolation of linear theory cannot match simulations  \citep{ColbergEtal2000, Peel2006}.

The halo model framework provides a phenomenological description of nonlinearities. The distribution of halo peculiar velocities ($\vect{v}$) depends on halo mass ($M_h$) and the environment, quantified by the local density contrast ($\tilde{\delta}$) smoothed over scale ($R$),
\begin{equation}
p(\vect{v}) \, \rmd \vect{v}= \frac{\iint p(\vect{v}|M_h,\tilde{\delta}) \, n(M_h|\tilde{\delta}) \, p(\tilde{\delta}) \, {\rm d} \tilde{\delta} \, {\rm d} M_h }
{\iint n(M_h|\tilde{\delta}) \, p(\tilde{\delta}) \, {\rm d} \tilde{\delta} \, {\rm d} M_h} \, \rmd \vect{v}\ ,
\label{eq:halomodel}
\end{equation}
where $p(\tilde{\delta})$ is the distribution of $\tilde{\delta}$, $n(M_h|\tilde{\delta})$ is the halo abundance of mass ($M_h$) at fixed $\tilde{\delta}$, and $p(\vect{v}|M_h,\tilde{\delta})$ is the conditional velocity distribution \citep{Sheth-Diaferio2001, HamanaEtal2005}. These $\tilde{\delta}$-dependent ingredients must be determined. The value $n(M_h|\tilde{\delta})$ follows directly from the conditional halo mass function \citep{HamanaEtal2005} or the halo bias function \citep{Sheth-Diaferio2001}. The conditional distribution $p(\vect{v}|M_h,\tilde{\delta})$ is typically assumed to be Gaussian, with variance ($\sigma_G^2$) empirically parameterized as a function of $M_h$ and $\tilde{\delta}$ and calibrated against simulations. Interestingly, even a crude approximation to $p(\tilde{\delta})$ appears sufficient for implementing Eq.~\ref{eq:halomodel}. Comparisons with simulations confirm the model’s validity. \citet{Sheth-Diaferio2001} asserted that the non‑Gaussianity, manifesting as long exponential‑like tails, arises primarily from integrating the Gaussian $p(\vect{v}|M_h,\tilde{\delta})$ over $\tilde{\delta}$.
Utilizing the model, \citet{BK2007, BK2008} demonstrated that the peculiar velocity distribution of halos, when combined with moments of the pairwise velocity, can potentially be an effective probe for constraining cosmological parameters. Interestingly, Eq.~\ref{eq:halomodel} was also adopted to compute the number density of direct-collapse black holes \citep{InayoshiEtal2015}.

Despite its apparent success, the halo model still suffers from limitations that impede its practical use. Notably, neither the choice of the smoothing scale, $R$, for defining $\tilde{\delta}$ nor the empirical $\sigma_G$-$\tilde{\delta}$ relation has a solid physical justification. Neither has been fully tested with simulations across different cosmological models. The model involves too many parameters that require calibration against simulations. It therefore seems desirable to find a simpler and more readily applicable formula.

We notice that the cumbersome halo model (Eq.~\ref{eq:halomodel}) essentially falls into the category of superstatistics proposed by \citet{BeckCohen2003}, which asserts that averaging the Boltzmann–Gibbs statistics over a distribution of dispersion can yield a family of heavy‑tailed statistics close to Tsallis statistics (sometimes termed $q$ statistics). These are particularly helpful for characterizing nonlinear systems typical of long‑range interactions and/or nontrivial correlations \citep{Tsallis2023}. Tsallis statistics are derived from the proposal of non‑extensive statistical mechanics, designed to generalize classical statistical mechanics beyond thermal equilibrium based on Tsallis entropy \citep{Tsallis1988}. In fact, Tsallis statistics have proven very useful for studying space plasmas \citep[e.g.,][]{PierrardLazar2010, LivadiotisMcComas2013}, although they were referred to as $\kappa$ distributions as early as \citet{Vasyliunas1968}. Over the decades, numerous astronomical applications have emerged, including matter distributions in halos \citep[e.g.,][]{HansenEtal2005, WojtakEtal2005, KronbergerEtal2006, BarnesEtal2007} and galaxies \citep[e.g.,][]{CardoneEtal2011, FrigerioEtal2015, SanchezEtal2021}, turbulence in the interstellar medium \citep[e.g.,][]{EsquivelLazarian2010, GonzalezEtal2018}, and the distribution of stellar rotational velocities \citep[][]{SoaresEtal2006, CarvalhoEtal2009, SilvaEtal2023}. \citet{LavagnoEtal1998} showed that the peculiar velocity distribution of a galaxy cluster sample can be well fitted by Tsallis statistics.

These considerations motivate us to systematically test Tsallis statistics using large suites of modern $N$‑body simulations to assess whether this framework can effectively capture the features of the halo velocity distribution and serve as a diagnostic tool for non‑Gaussianity and its evolution. This paper is structured as follows. In Sect. 2, we provide a brief introduction to non‑extensive statistical mechanics and Tsallis statistics. Sections 3 and 4 present the fitting results of the Tsallis statistics to the peculiar velocity distribution functions of halos in N-body simulations from two different simulation projects. Section 5 discusses the interpretation within the superstatistics framework and possible avenues for improving the halo model. We provide a summary of our study in Sect. 6.

\section{Tsallis statistics}

\subsection{Tsallis entropy}

A pair of functions plays a key role in the theory: 
\begin{equation}
\begin{aligned}
\ln_q x & \equiv (x^{1-q}-1)/(1-q)\ , \quad  \text{for } x >0\ ;\\
e_q(x) & \equiv \begin{cases} \left[ 1+(1-q)x\right]^{1/(1-q)}  & \text{if } 1+(1-q)x >0  \\
						0  &\text{otherwise}.
			\end{cases}
\end{aligned}
\end{equation} 
One can verify that $e_q[\ln_q(x)]=\ln_q[e_q(x)]=x$, and in the limit of $q\to 1$, $\ln_q \to \ln$, and $e_q \to e$. 
The essence of non-extensive statistical mechanics lies in the introduction of Tsallis entropy \citep{Tsallis1988}:
\begin{equation}
S_q = - k_B\sum_i p_i^q \ln_q p_i= -k_B \frac{1-\sum_i p_i^q}{1-q}\ ,
\end{equation}
where $k_B$ is the Boltzmann constant, and the sum runs over all possible microstates, each with probability $p_i$. When $q=1$, this reduces to the classical entropy, $S=-k_B \sum_i p_i \ln p_i$.

The Tsallis entropy is non-extensive. For a composite system, $A\cup B$, consisting of two independent subsystems, $A$ and $B$ (i.e., where their microstate probabilities factorize),
$S_q(A\cup B) = S_q(A)+S_q(B)+[(1-q)/k_B]S_q(A) S_q(B)$, in contrast to the classical relation, $S(A\cup B) = S(A)+S(B)$.

\subsection{Tsallis distribution about velocity}

For a system consisting of $N$ self-gravitating particles of mass $m$, the distribution of 3D velocity ${\bf v}=(v_x, v_y, v_z)$ is modeled by
\begin{equation}
\begin{aligned}
& p(\vect{v}) \rmd \vect{v}=  C_3 \left( \frac{\abs{1-q_3}}{ \pi \sigma_3^2} \right)^{3/2} \left[ 1-(1-q_3) \frac{v^2}{\sigma_3^2} \right]^{1/(1-q_3)} \rmd \vect{v}\ , \\
& C_3 = \begin{cases}
\displaystyle
\Gamma\left(\frac{1}{1-q_3} + \frac{5}{2}\right)\big/ \Gamma\left( \frac{1}{1-q_3}+1\right) & \text{for } q_3< 1 \ , \\
\displaystyle
\Gamma\left( \frac{1}{q_3-1}\right) \big/ \Gamma\left( \frac{1}{q_3-1}- \frac{3}{2} \right) & \text{for }   1< q_3 < \frac{7}{5}\ . 
\end{cases}
\end{aligned}
\label{eq:pv3d}
\end{equation}
Here, $v=\abs{\vect{v}}$, and $\Gamma$ denotes the gamma function \citep{SilvaEtal1998, LimaEtal2002}. For $q_3<1$, the speed has a cutoff at $v_{\max}= \sigma_3 /\sqrt{1-q_3}$. Additionally, the mean kinetic energy of the particles, $\avg{\epsilon} \propto \avg{v^2}=3\sigma^2_3/(7-5q_3)$, must be positive, which requires $q_3< 7/5$.

Several methods exist to derive the velocity distribution. One approach maximizes the Tsallis entropy via Lagrange multipliers, subject to a constraint on the $q$ expectation of kinetic energy, $ \avg{\epsilon}_q = \int (mv^2/2) p^q(\vect{v}) \rmd \vect{v}/ \int p^q( \vect{v}) \rmd \vect{v}$ \citep{CuradoTsallis1991}, where the function $p^q(x)/\int p^q(x)\rmd x$ is known as the escort probability. A kinetic approach instead assumes
\begin{equation}
\ln_q p(\vect{v})  = \ln_q p(v_x) +  \ln_q p(v_y) +  \ln_q p(v_z)
\label{eq:facto}
\end{equation}
instead of the usual factorization $p(\vect{v})= p(v_x)p(v_y)p(v_z)$ and follows the classical Maxwell's approach to obtain Eq.~\ref{eq:pv3d} \citep{SilvaEtal1998}. Moreover, it has been shown that the Boltzmann equation for the system admits a generalized $H$-theorem based on Tsallis entropy. Consequently, the system evolves irreversibly toward a collisional equilibrium state given by Eq.~\ref{eq:pv3d} \citep{LimaEtal2001, RamosEtal2004}.

\subsection{Marginalization and factorization}

The 1D velocity distribution derived from Eq.~\ref{eq:facto} is
\begin{equation}
\begin{aligned}
& p(v_1) \rmd v_1=C_1 \left( \frac{\abs{1-q_1}}{\pi \sigma_1^2}\right)^{1/2}\left[ 1-(1-q_1) \frac{v_1^2}{\sigma_1^2} \right]^{1/(1-q_1)}\rmd v_1\ , \\
&C_1 = 
\begin{cases}
\Gamma\left(\frac{1}{1-q_1} + \frac{3}{2}\right) \big/ \Gamma\left( \frac{1}{1-q_1} +1 \right) & \text{for } q_1< 1 \ , \\
\Gamma\left( \frac{1}{q_1-1}\right)  \big/  \Gamma\left( \frac{1}{q_1-1} -\frac{1}{2} \right) & \text{for } 1< q_1 < \frac{5}{3}\ .
\end{cases}
\end{aligned}
\label{eq:pv1d}
\end{equation}
For $q_1<1$, the distribution is restricted to $\abs{v_1} \leq \sigma_1/\sqrt{1-q_1}$. The condition, $q_1<5/3$, follows from the requirement that $\avg{v_1^2}=\int p_(v_1)v_1^2\rmd v_1= \sigma_1^2/(5-3q_1) > 0$. This distribution also maximizes Tsallis entropy for a 1D system.

The 1D distribution obtained by marginalizing the 3D distribution -- for example, $\tilde{p}(v_x)=\iint p(\vect{v})\rmd v_y \rmd v_z$ -- does not exactly match Eq.~\ref{eq:pv1d} since $q_1\neq q_3$. Direct integration yields
\begin{equation}
\begin{aligned}
& \tilde{p}(v_x) = \widetilde{C}_1\left( \frac{\abs{1-q_3}}{\pi \sigma_3^2}\right)^{1/2} \left[  1-(1-q_3) \frac{v_x^2}{\sigma_3^2} \right]^{1/(1-q_3)+1}\ , \\
&\widetilde{C}_1 = 
\begin{cases}
\Gamma\left(\frac{1}{1-q_3} + \frac{5}{2}\right) \big/ \Gamma\left( \frac{1}{1-q_3} +2 \right) & \text{for } q_3< 1 \ , \\
\Gamma\left( \frac{1}{q_3-1}-1\right)  \big/  \Gamma\left( \frac{1}{q_3-1} -\frac{3}{2} \right) & \text{for } 1<q_3<\dfrac{7}{5}\ .
\end{cases}
\end{aligned}
\label{eq:pv1d_mg}
\end{equation}
The marginalized distribution has two notable properties relative to Eq.~\ref{eq:pv3d}:
\begin{itemize}
\item For $q_3<1$,  $\abs{v_x}_{max}=\abs{v}_{max}= \sigma_3/\sqrt{1-q_3}$. 
\item  Moreover, $q_3<7/5$, because $\avg{v_x^2}= \sigma_3^2/(7-5q_3)=\avg{v^2}/3>0$. 
\end{itemize}
Hence, to rewrite Eq.~\ref{eq:pv1d_mg} in the form of Eq.~\ref{eq:pv1d}, we require that $q_1=1/(2-q_3)$ and ${\sigma_1}^2=\sigma_3^2 /(2-q_3)$.

The above result appears to pose a puzzling challenge to probability theory, since one would expect $q_1=q_3$. \citet{SilvaEtal1998} argued that the marginalization procedure in the context of non-extensive mechanics should be modified as follows:
\begin{equation}
p(v_1, v_2, \ldots, v_m)=\frac{\int p^\alpha (v_1, v_2, \ldots, v_m, \ldots, v_n) \rmd v_{m+1} \ldots \rmd v_n}{\int p^\alpha (v_1, v_2, \ldots, v_m, \ldots, v_n) \rmd v_1 \ldots \rmd v_n}, 
\end{equation}
with $\alpha = 1- (1-q)(n-m)/2$. For our problem, $m=1$ and $n=3$ gives $\alpha=q$, which in fact corresponds to marginalizing the escort probability function over $v_y$ and $v_z$. 

Instead of modifying the concept of marginalization, another feasible solution to the issue is to allow a dimension-dependent $q_d$ in the factorization of Eq.~\ref{eq:facto}. This ad hoc assumption is expressed as
\begin{equation}
\ln_{q_3} p(\vect{v})=\ln_{q_1} p(v_x) +  \ln_{q_1} p(v_y) +  \ln_{q_1} p(v_z), 
\label{eq:newfacto}
\end{equation}
with $q_3=2-1/q_1$ to account for the change in dimension. Clearly, with this treatment, one can comfortably follow the classical Maxwell derivation and obtain consistent distribution functions across dimensions. 

\subsection{$\kappa$-notation}

It has been argued that dimensional dependence reflects a dependence on the degrees of freedom, and that the $d$-dimensional distribution can be properly expressed using a different $\kappa$ notation by introducing an invariant index ($\kappa_0$) independent of $d$ \citep{LivadiotisMcComas2011, Livadiotis2015, Livadiotis2015_JGR}. Accordingly, in this work, the $\kappa$ notation was adopted for both our numerical and theoretical analyses to avoid any confusion that might arise from dimension dependence.

Explicitly, 
\begin{equation}
\begin{aligned}
&p_{\kappa_0}(\vect{v})= \frac{C_{\kappa_0} }{ \left( \pi \abs{\kappa_0} \sigma_{\kappa_0}^2\right)^{d/2}}\left[ 1+\frac{1}{\kappa_0} \frac{v^2}{\sigma_{\kappa_0}^2} \right]^{-\kappa_0-1-d/2}\ , \\
&C_{\kappa_0}   = \begin{cases}
\Gamma\left( \abs{\kappa_0}  \right)  \big/  \Gamma\left( \abs{\kappa_0}-d/2  \right) & \text{if } \kappa_0 < -(1+d/2)\ ,
\\
\Gamma\left(\kappa_0+1+d/2 \right) \big/ \Gamma\left( \kappa_0 +1 \right) & \text{if } \kappa_0 > 0 \ ,
\end{cases}
\end{aligned}
\label{eq:kappa}
\end{equation}
with $\kappa_0= 1/(q-1)-(1+d/2)$ and $\sigma_{\kappa_0}^2=\sigma_q^2/[1-(q-1)(1+d/2)]$ \footnote{For $\kappa_0 <0$, a meaningful $\avg{v_d^2}$ requires $\kappa_0 <-d/2$ \citep{Livadiotis2015_JGR}. In the $q$ notation  (Eqs.~\ref{eq:pv3d} and ~\ref{eq:pv1d}), $p(\vect{v}) \propto (1-Av^2/\sigma^2)^{1/A}$ with $A$ and $1/A$ having the same sign. However, if $\kappa_0\in (-1-d/2, -d/2)$,  then $p_{\kappa_0}\propto (1-\abs{\kappa_0}^{-1}v^2/\sigma^2)^{\abs{\kappa_0}-1-d/2}$. Because $\abs{\kappa_0}^{-1}>0$ and $\abs{\kappa_0} -1-d/2 <0$, recasting $p_\kappa$  into the $q$ notation is impossible. Hence, we restrict $\kappa_0 < -(1+d/2)$ for consistency. In contrast, Eq.~\ref{eq:pv1d_mg} avoids this issue, since its $q<7/5$.}. 
When $\kappa_0<-1-d/2$, the velocity cutoff  is $\sigma_{\kappa_0}\sqrt{\abs{\kappa_0}}$. As $\kappa_0 \to +\infty$, $p_{\kappa_0}$ tends to a Gaussian.

Odd velocity moments vanish due to symmetry: $p_{\kappa_0}(-\vect{v}) =p_{\kappa_0}(\vect{v})$. Even moments exist for $\kappa_0>n-1$ (equivalently, $q<1+1/(n+d/2)$) and are given by
\begin{equation}
\avg{\vect{v}^{2n}} = \frac{\Gamma(n+d/2)\Gamma(\kappa_0+1-n)}{\Gamma(d/2)\Gamma(\kappa_0+1)}
(\kappa_0 \sigma_{\kappa_0}^2)^n\ ,
\label{eq:mom}
\end{equation}
with $n=1,2,3\ldots$. For a Gaussian, $\avg{\vect{v}^{2n}}_G=(2\sigma_G^2)^n\Gamma(n+d/2)/\Gamma(d/2)$. In practice, the second and fourth moments provide a quick estimation of $(\kappa_0, \sigma_{\kappa_0})$ as initial fitting parameters.

\section{Halo velocity distribution in the Quijote $N$-body simulation}
\label{sec:sim}

\subsection{Data and methodology}

\begin{table*}[h!]
\caption{Cosmological parameters of the N-body simulations.}
\label{tab1}
\centering
\begin{tabular}{c | c c c c c c c c}
\hline\hline
Label  & $\Omega_m$ & $\Omega_b$ & $\Omega_\nu$ & $h$ & $n_s$ & $\sigma_8$ & $w_0$ & $w_a$  \\
\hline
Quijote Fid. HR & 0.3175 & 0.049 & 0 & 0.6711 & 0.9624 & 0.834 & -1.0 & 0.0   \\
\hline
Mira-Titan M000 & 0.2200 & 0.04479 & 0 & 0.7100 & 0.9630 & 0.8 & -1.0 & 0.0  \\
Mira-Titan M001 & 0.3276 & 0.05945 & 0 & 0.6167  & 0.9611 & 0.8778 & -0.7 & 0.6722  \\
Mira-Titan M010 & 0.1718 & 0.03649 & 0 & 0.7833 & 0.9389 & 0.7222 & -1.3 & -0.5222  \\
\hline                             
\end{tabular}
\end{table*}

This work used 100 high‑resolution realizations of the fiducial Lambda cold dark matter model ($\Lambda$CDM) from the Quijote $N$‑body simulations \citep[Quijote Fid. HR;][]{Quijote2020}\footnote{\url{https://quijote-simulations.readthedocs.io/en/latest/types.html}}. The fiducial cosmological parameters follow \citet{Planck2018}, as listed in Table~\ref{tab1}. Each realization simulated $1024^3$ particles of mass $8.207 \times 10^{10}\msun$ in a $1\ h^{-1}\ \mathrm{Gpc}$ cubic box.

In total, 300 halo catalogs were constructed using the friends‑of‑friends algorithm with a linking length parameter $b=0.2$, applied to the 100 realizations at three redshifts $z=0,1,2$. The halos contain at least $20$ member particles, corresponding to a minimum halo mass of $M_h \geq 1.6414\times 10^{12}\msun$. On average, each realization contained about 3, 2.8, and 1.3 million halos at $z=0,1,2$ with mean masses of $9.95\times 10^{12}\msun$, $5.94\times 10^{12}\msun$, and $4.11\times 10^{12}\msun$ respectively. 

We adopted Eq.~\ref{eq:kappa} as the fitting template. The mean velocity $\avg{\vect{v}}$ was not treated as a free parameter. In fact, we find that $\abs{\avg{\vect{v}}}/\avg{\vect{v}^2}^{1/2} \lesssim 0.25\%$ for all halo samples, and including $\avg{\vect{v}}$ changes the results by less than $1\%$. Since the halo samples are free of selection effects and observational uncertainties, we used a simple unweighted maximum-likelihood estimate (MLE). For a sample of N halos, the likelihood is
\begin{equation}
\mathcal{L}(\kappa_0, \sigma_\kappa) = \prod_{i=1}^N p_{\kappa_0} (\vect{v}_i; \kappa_0, \sigma_\kappa)\ , 
\label{eq:mle}
\end{equation}
and we obtained the parameters $(\kappa_0,\, \sigma_{\kappa_0})$ by numerically maximizing $\ln \mathcal{L}$. 

When the MLE is difficult to apply to a very large number of halos or when only the binned velocity data are available, a coarse‑grained version of Eq.~\ref{eq:mle} can be employed. Let $P_j$ be the number of halos with peculiar velocities falling into the $j$th bin centered at $\vect{v}_j$. The likelihood is then approximated by
\begin{equation}
\ln \mathcal{L} \approx \sum_j P_j \ln p_{\kappa_0}(\vect{v}_j;\kappa_0, \sigma_\kappa)\ .
\label{eq:histMLE}
\end{equation}
Naturally, the accuracy of this approximate MLE depends on the histogram bin width.

\subsection{Distribution functions}

For each realization, the parameters were estimated via MLE using Eq.~\ref{eq:mle}. For comparison, the probability density functions (PDFs) of $v_x$, $v_y$, $v_z$, and $v=\abs{\vect{v}}$ for the simulated halos ($p_{\rm sim}$) were obtained from histograms with a bin size of $5\, \kms$. The best-fit PDFs $p_{\rm fit}$ were smoothed to compensate for the binning effect, and relative differences $p_{\rm sim}/p_{\rm fit}-1$ were plotted subsequently to assess the goodness of fit.

\begin{figure*}[h!]
   \centering
   \resizebox{\hsize}{!}{ \includegraphics{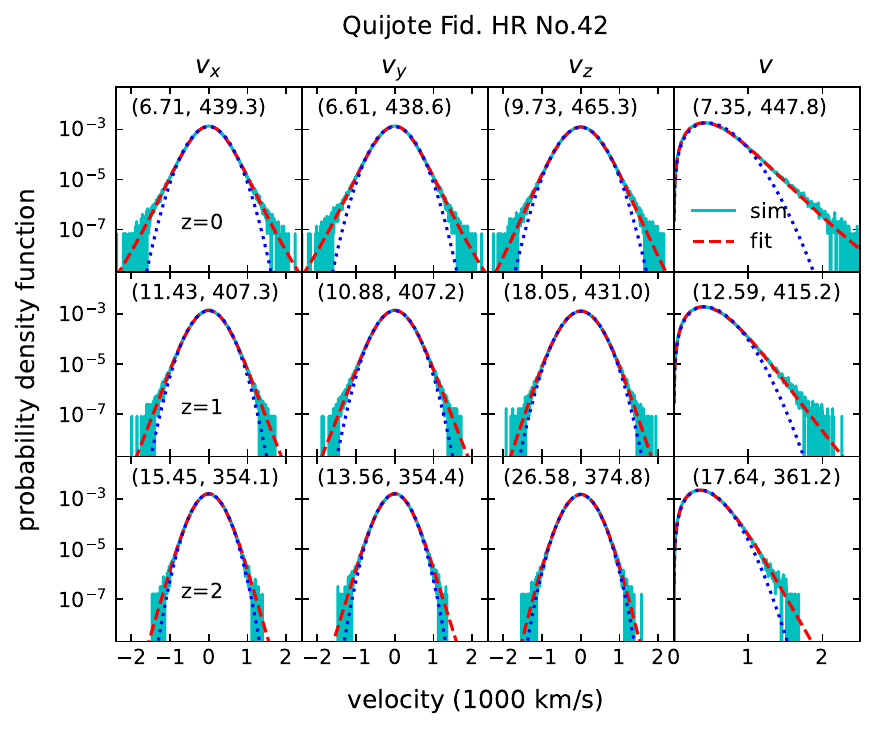} \includegraphics{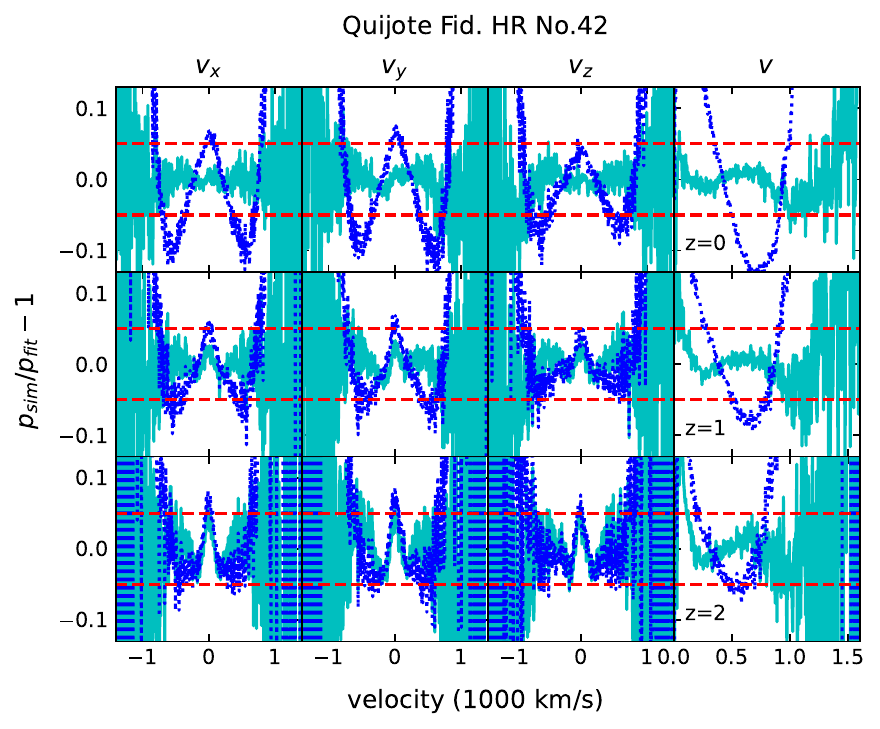}}
      \caption{Left panel: Halo velocity distribution functions (solid cyan lines) from realization No.~42 of the Quijote fiducial high-resolution simulation at redshifts $z = 0, 1, 2$. The dashed red lines represent the best fits derived with Eq.~\ref{eq:kappa} with the corresponding $(\kappa_0, \sigma_{\kappa_0})$ parameters listed in each subpanel. The dotted blue lines represent the best-fit Gaussian models. Right panel: Relative differences between the simulation and the best-fit Tsallis (cyan lines) and Gaussian (blue dotted lines) models. The horizontal dashed red lines indicate the $5\%$ precision level. The velocity bin width is $5\ \kms$.}
         \label{fig:onefit}
\end{figure*}

Figure~\ref{fig:onefit} lists our results from a randomly selected realization (No. 42 of 100). The Tsallis statistics provide a much better fit to the simulation than the Gaussian model. A precision of $5\%$ is generally achieved for $\abs{\vect{v}} \lesssim 1000\ \kms$, with the highest accuracy at $z=0$. As redshift decreases toward zero, $\kappa_0$ declines, indicating that the $N$‑body system moves further from equilibrium and Gaussianity over time. Thus, gravitational interaction drives the system from an extensive state toward a non‑extensive one.

Velocity histograms from individual realizations are too noisy, particularly at high speeds. We therefore combined the halo catalogs from all 100 realizations into a single stacked sample. Because of its large size, we used the coarse‑grained MLE (Eq.~\ref{eq:histMLE}) to ensure that the computation remains feasible on a standard desktop computer. Numerical tests indicate that with a bin size of
$1\ \kms$, the relative differences from the full MLE (Eq.~\ref{eq:mle}) are typically well below $1\%$.

\begin{figure*}[h!]
   \centering
   \resizebox{\hsize}{!}{ \includegraphics{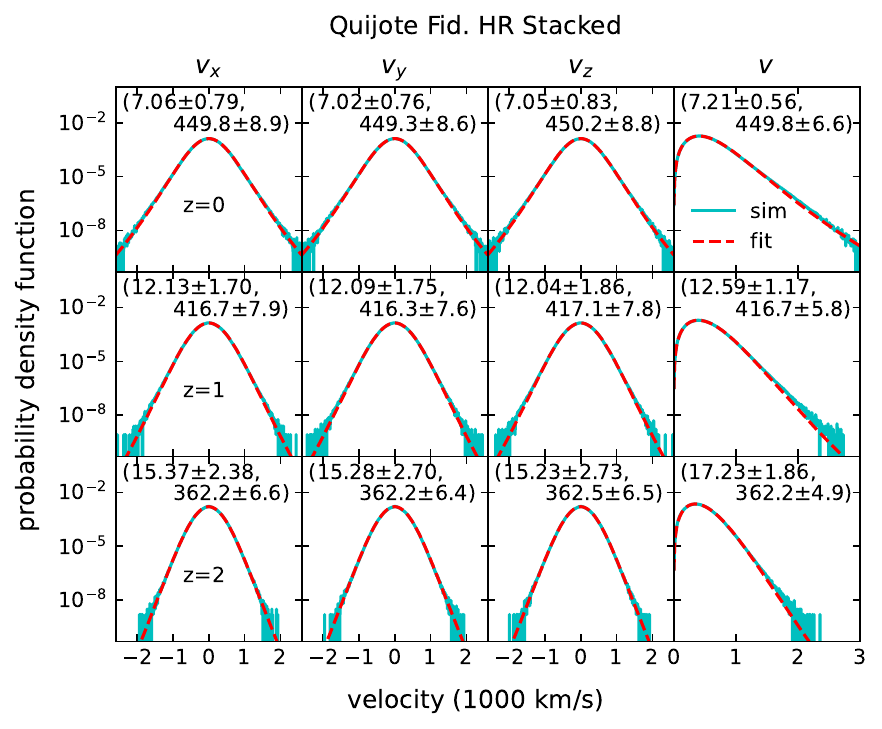}\includegraphics{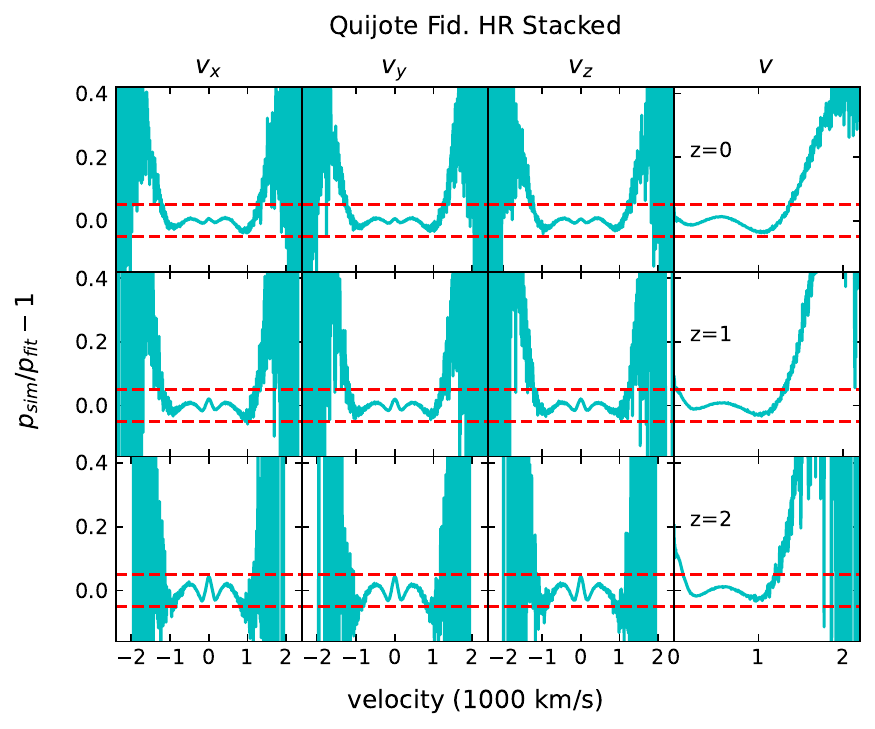}}
      \caption{Left panel: $p_{\rm sim}$ (cyan solid) and $p_{\rm fit}$ (red dashed) for the stacked halo samples at $z = 0, 1, 2$. The parenthetical numbers represent the best fits $(\kappa_0, \sigma_{\kappa_0})$. Right panel: Relative differences between $p_{\rm sim}$ and $p_{\rm fit}$. The horizontal dashed red lines indicate $5\%$ precision. The velocity bin width is $5\ \kms$.}
         \label{fig:stackedfit}
 \end{figure*}

Velocity distribution functions of the stacked halo sample and the best‑fit Tsallis models are presented in Fig.~\ref{fig:stackedfit}. From Figs.~\ref{fig:onefit} and~\ref{fig:stackedfit}, we identify two regimes where Eq.~\ref{eq:kappa} is not entirely satisfactory.
\begin{itemize}
\item High speed tails for $\abs{\vect{v}} \gtrsim 1000 \ \kms$.  Although the Tsallis function is known to typically exhibit power‑law‑like heavy tails, the best‑fit Tsallis model underestimates the fraction of high‑speed halos. At $v = 2000$km/s and $z=0$, the difference between $p_{\rm sim}$ and $p_{\rm fit}$ is already $\sim 50\%$, and the disagreement in the high‑speed tails becomes more severe at higher redshifts. 
\item Behavior near $v =0$ for $z\geq 1$.   The difference in $p(v)$ between the best-fit Tsallis model and the simulation near $v=0$ becomes moderately large when $z \geq 1$, reaching $10\sim 20\%$.
\end{itemize}

Weighting high‑speed halos in the MLE could improve the tail fit but at the cost of reduced accuracy at low speeds. Achieving percent‑level precision across the full velocity range likely requires additional modeling. Nevertheless, the Tsallis distribution approximation is likely acceptable for many applications, since the 1$\sigma$ uncertainties in the tails of the PDFs' -- where $p(v) < 4/N$ (roughly $v\gtrsim 10^3$km/s) -- are as large as $\approx 0.5\sqrt{p(v)/N}$ in any case. Within these uncertainties, the simple MLE‑fitted Tsallis distributions are consistent with the simulation results. Any additional ingredients would therefore have only a minor effect.

From these results, we also find that, for an individual halo sample, the four sets of parameters estimated from the three 1D components $v_x$, $v_y$, and $v_z$ and the 3D vector $\vect{v}=(v_x, v_y, v_z)$ are by no means identical. There is considerable stochasticity, as indicated by the $\kappa_0$ values shown in Fig.~\ref{fig:onefit}. Differences in $\kappa_0 \sigma_{\kappa_0}^2$, which dominate the velocity moments, can also exceed $\sim 50\%$. This could be a significant source of statistical uncertainty for applications based on 1D velocity data. Of course, the anisotropy is greatly suppressed in the stacked sample (Fig.~\ref{fig:stackedfit}), and statistical isotropy can be recovered by averaging over a large volume if the ergodicity assumption holds.

\subsection{Moments and correlated components}

\begin{figure}[h!]
   \centering
   \resizebox{\hsize}{!}{ \includegraphics{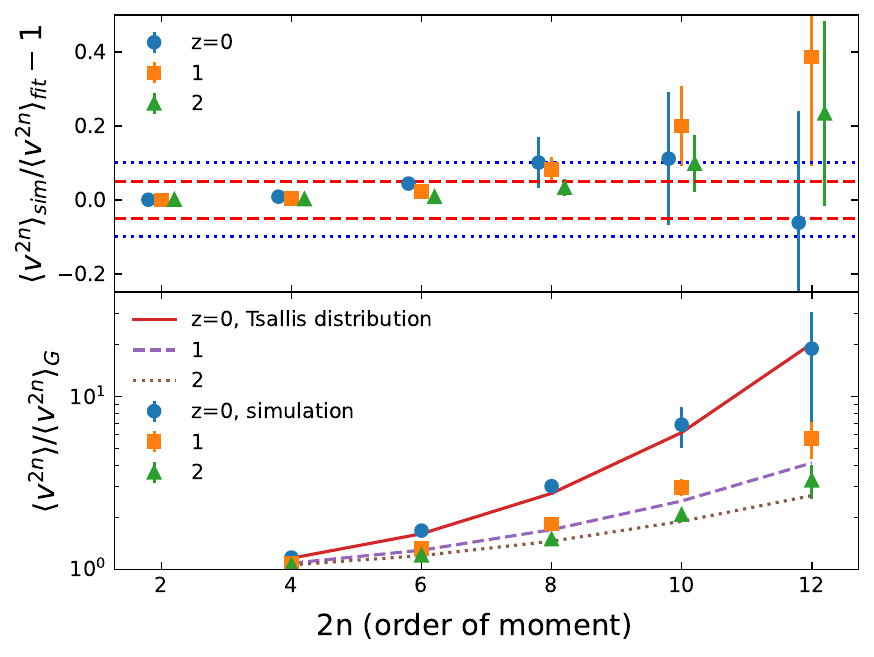}}
      \caption{Comparison of halo velocity moments from the simulations with the best‑fit Tsallis predictions. In the top panel, data points at different redshifts slightly offset horizontally for clarity. The horizontal dashed red and dotted blue lines indicate $5\%$ and $10\%$ deviation, respectively. In the bottom panel, the Gaussian model, $\langle v^{2n} \rangle_G$, has the same variance as the simulation. The error bars indicate the standard deviations across 100 realizations.}
         \label{fig:mom}
\end{figure}

\begin{figure}[h!]
   \centering
   \resizebox{\hsize}{!}{ \includegraphics{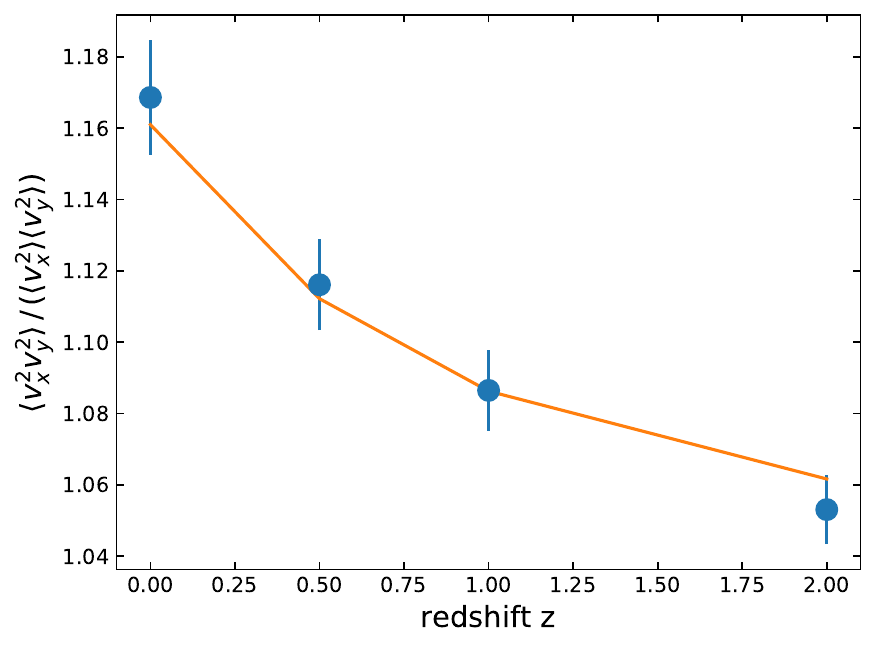}}
      \caption{Quadratic correlation between the velocity components of halos. The symbols represent measurements from the simulation, and the solid line represents the prediction of the best-fit Tsallis model.}
         \label{fig:xyz}
 \end{figure}

A comparison of moments up to 12th order \footnote{All MLE-fitted $\kappa_0 > 5.5$.} shows that the Tsallis approximation (Eq.~\ref{eq:kappa}) is practically useful (Fig.~\ref{fig:mom}). It accurately captures the non‑Gaussianity in $p(\vect{v})$ at the moment level. Moreover, the Tsallis model predicts statistical dependence among velocity components, i.e., $p(\vect{v})\neq p(v_x)p(v_y)p(v_z)$. Consider $v_x$ and $v_y$. Although the linear correlation $\avg{v_x v_y}=0$, it does not imply independence. Isotropy gives $\avg{v_x^2 v_y^2}=\avg{v^4}/15$ and $\avg{v_x^2}=\avg{v_y^2}=\avg{v^2}/3$. The Gaussian model has $\avg{v^4}/\avg{v^2}^2=5/3$, leading to $\avg{v_x^2 v_y^2} = \avg{v_x^2}\avg{v_y^2}$. In contrast, the Tsallis model yields 
\begin{equation}
\frac{\avg{v_x^2 v_y^2}}{\avg{v_x^2}\avg{v_y^2}} = \frac{\kappa_0}{\kappa_0-1} \ ,
\end{equation}
which agrees excellently with our simulations (Fig.~\ref{fig:xyz}). In fact, one can easily show that
\begin{equation}
\avg{v_x^{2n_1} v_y^{2n_2} v_z^{2n_3}}=\avg{\vect{v}^{2n}}\frac{\Gamma(n_1+1/2)\Gamma(n_2+1/2)\Gamma(n_3+1/2)}{2\pi \Gamma(n+3/2)}\ ,
\end{equation}
with $n=n_1+n_2+n_3$. The correlation structure is fully governed by the velocity moments. These nontrivial higher-order correlations may affect the modeling of redshift-space distortions and the velocity ellipsoid of galaxy clusters and could serve as additional diagnostics of non-Gaussianity in velocity space.

\subsection{Mass dependence}
\label{sec:mass}

\begin{figure*}[h!]
   \centering
   \resizebox{\hsize}{!}{ \includegraphics{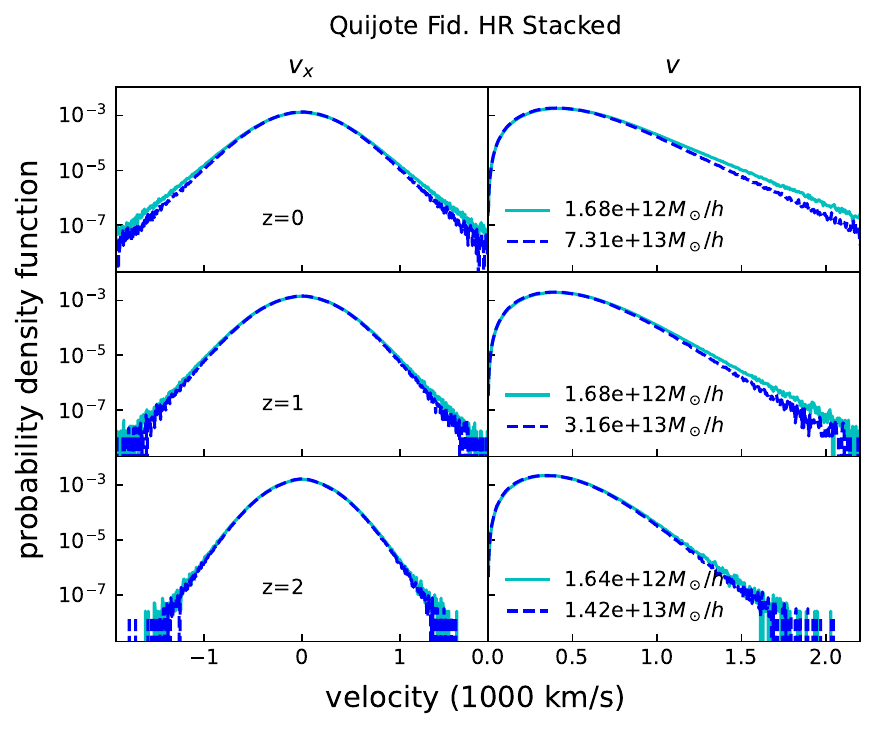}\includegraphics{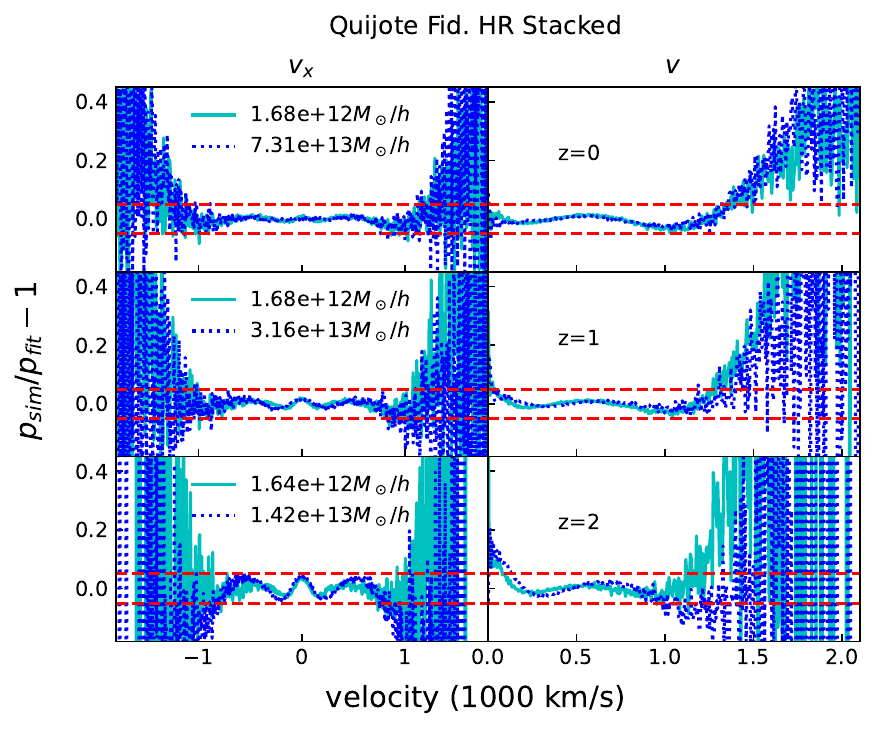}}
      \caption{Left: $p_{\rm sim}$ (solid cyan lines) for the stacked halo samples in the lightest and heaviest mass bins at $z=0,1,2$. Right: Relative differences $p_{\rm sim}/p_{\rm fit}-1$ with $p_{\rm fit}$ from the best-fit Tsallis models. The horizontal dashed red lines indicate $5\%$ precision. Mean masses are indicated in the legends. The bin width is $5\ \kms$.}
         \label{fig:substacked}
 \end{figure*}
 
Having established that the Tsallis model captures the non-Gaussianity and nontrivial correlations of halo velocities, we now examine how these statistical properties depend on halo mass. Halo samples at $z=0,1,2$ were divided into $13$, $13$, and $10$ distinct mass subsamples, respectively. At a given epoch, each mass subsample contains roughly the same number of halos. Stacked subsamples were produced by merging those generated from different realizations belonging to the same mass bin. The Tsallis model parameters (Eq.~\ref{eq:kappa}) for the stacked mass subsamples were estimated using the coarse‑grained MLE (Eq.~\ref{eq:histMLE}) applied to velocity histograms with a resolution of $1\ \kms$. The PDFs of two stacked subsamples — the lightest and the heaviest — at the three redshifts are shown in Fig.~\ref{fig:substacked}.

\begin{figure}[h!]
   \centering
   \resizebox{\hsize}{!}{ \includegraphics{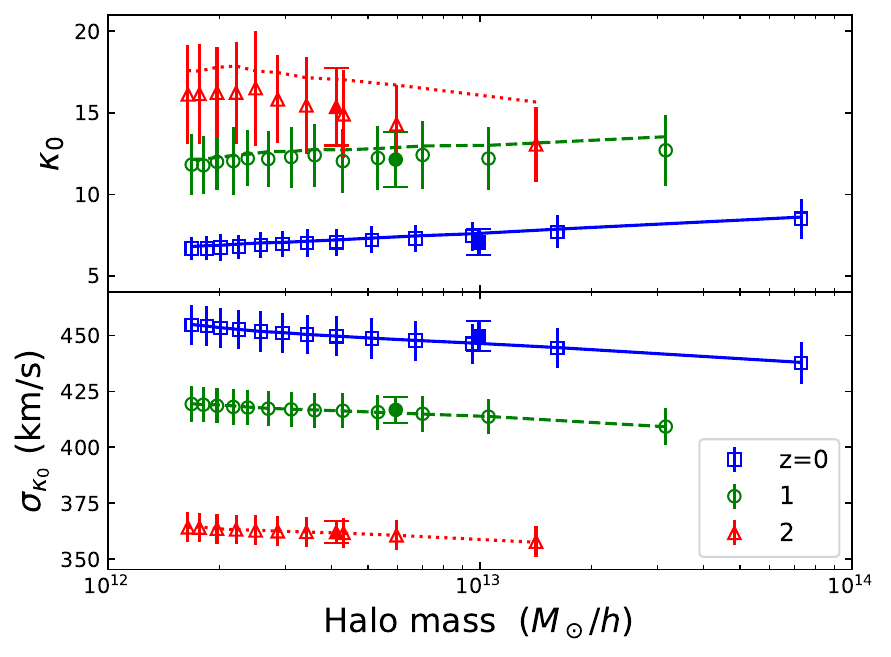}}
      \caption{Mass dependence of $(\kappa_0, \  \sigma_{\kappa_0})$. The unfilled symbols represent the estimated measurements derived from the PDFs of $v_x$ for the mass subsamples, while the filled symbols with the capped error bars correspond to the full catalogs. The lines represent the results obtained from $p(\vect{v})$.}
         \label{fig:pars_mass}
 \end{figure}

A major conclusion is that the Tsallis distribution consistently fits the $p(\vect{v})$ of halos across all mass bins very well. The goodness of fit at high‑speed tails actually appears even better for massive halos at high redshifts. Differences among the $p(\vect{v})$ curves are fully captured by variations in $(\kappa_0,\ \sigma_{\kappa_0})$, whose mass dependence is shown in Fig.~\ref{fig:pars_mass}. One can easily recognize that the mass dependence of $p(\vect{v})$ is marginal but discernible. Generally speaking, the peculiar velocity distribution functions of low‑mass halos develop heavier high‑speed tails (larger $\sigma_{\kappa_0}$) and exhibit stronger non‑Gaussianity (smaller $\kappa_0$ than those of massive halos, except at $z=2$, where $\kappa_0$ decreases at the high‑mass end). It appears that the velocity bias of halos is more complex than that suggested using the peak‑background split method.

We also find that at $z=2$, the $\kappa_0$ derived from $p(\vect{v})$ is systematically larger than that derived from $p(v_x)$, though the two are consistent within $1\sigma$. This, together with the unusual mass dependence of $\kappa_0$ at this redshift, likely reflects numerical artifacts: the Tsallis model loses accuracy at high $z$ and may become inadequate for such an analysis.

Halos and their peculiar velocities are biased tracers of underlying dark matter \citep[e.g.,][]{BardeenEtal1986, CooraySheth2002}. The peak‑background split approach suggests that velocity bias depends only on halo mass \citep[e.g.,][]{BardeenEtal1986, DesjacSheth2010}. Measurements from the power spectrum show that the velocity bias of halos deviates from unity by only a small amount \citep{ChenEtal2018}, likely because halos are peaks of the non‑Gaussian density field \citep{Zhang2018}. The theory assumes a deterministic velocity bias. While $\sigma_{\kappa_0}$ may vary with mass, $\kappa_0$ should remain essentially constant. Figure~\ref{fig:pars_mass} shows that at $z=1$, $\kappa_0$ is nearly constant across mass bins, but at $z=0$, the most massive halos have a notably higher $\kappa_0$ than the least massive ones. This indicates that stochasticity in velocity bias becomes nonnegligible at the late stages of gravitational evolution.  

\section{Validation with the Mira-Titan suite}

\begin{table*}[h!]
\caption{Characteristic parameters of Mira-Titan halo catalogs.}
\label{tab2}
\centering
\begin{tabular}{c | c | c c c | c c c }
\hline
Label  & particle mass $(10^{9}\msun)$ & \multicolumn{3}{| c |}{Halo Numbers $(10^7)$} & \multicolumn{3}{| c }{Mean halo mass $(10^{12} \msun)$}   \\
          &          & $z=0$ & $z=1$ & $z=2$ & $z=0$ & $z=1$ & $z=2$ \\
\hline
Mira-Titan M000 & 7.434 & 6.5 & 5.9 &  4.2& 1.312 & 0.807 & 0.550   \\
Mira-Titan M001 & 7.121 & 6.6  & 6.8 &  6.0 & 1.145 &  0.927 & 0.667 \\
Mira-Titan M010 & 7.852 & 6.2 & 4.9 & 2.6   & 1.181 & 0.709 & 0.471 \\
\hline                             
\end{tabular}
\end{table*}

It is conceivable that the accurate fit of the Tsallis model to the Quijote Fid. HR simulation data is simply a coincidence resulting from the specific cosmological parameters used. Hence, it is necessary to verify the model with simulations of different cosmological parameters. Given that other simulation data from the Quijote project -- compared to the Quijote Fid. HR suite -- either have mass resolutions nearly an order of magnitude lower or lack sufficient realizations to provide a comparable number of dark matter halos, we instead use the Mira-Titan Universe suite of the HACC simulation project \citep{HabibEtal2016, HeitmannEtal2016, HeitmannEtal2019} \footnote{\url{https://cosmology.alcf.anl.gov}}. This choice also avoids coincidences arising from the use of the same algorithms, computers, and simulation personnel (though the probability of such coincidences is expected to be negligibly small). The three realizations selected from the Mira-Titan Universe are based on distinct cosmological parameters (Table~\ref{tab1}). Each realization is composed of $3200^3$ particles running in a cubic box of size $2100^3\textrm{Mpc}^3$. Outputs at $z=0, 1, 2$ are primarily used for validation. Halos are identified by the frends-of-friends algorithm with a linking length $b=0.168$. The minimal number of member particles is 20. The numbers of halos in the three Mira-Titan simulations are approximately $1/5$ of the total numbers of halos summed over the 100 realizations of the Quijote Fid. HR simulation at the corresponding redshift, and mean halo masses are also lower (Table~\ref{tab2}).

\begin{figure*}[h!]
\centering
   \resizebox{\hsize}{!}{ \includegraphics{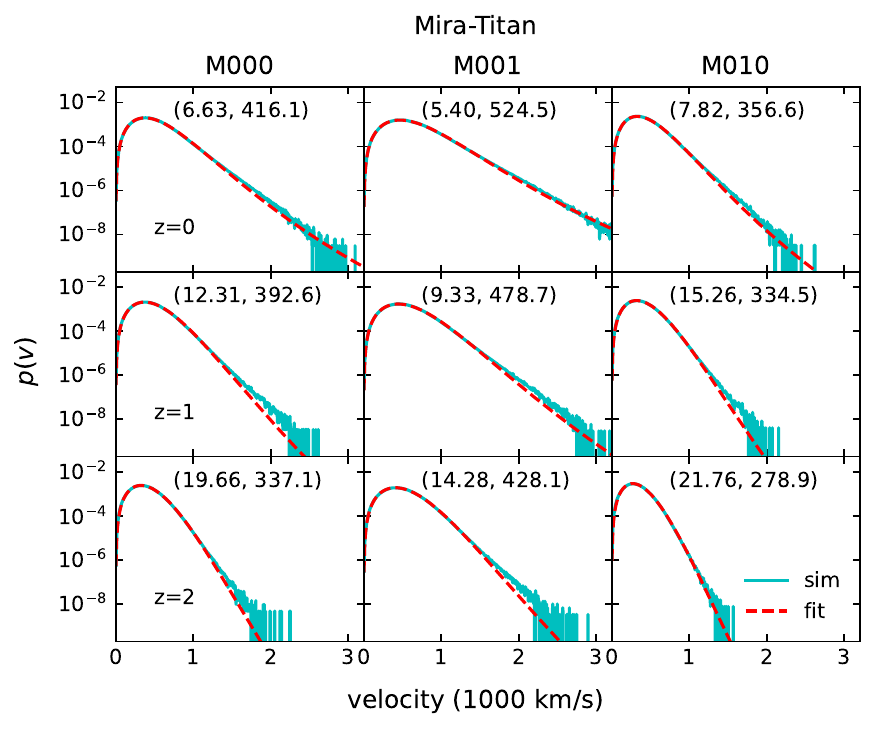}\includegraphics{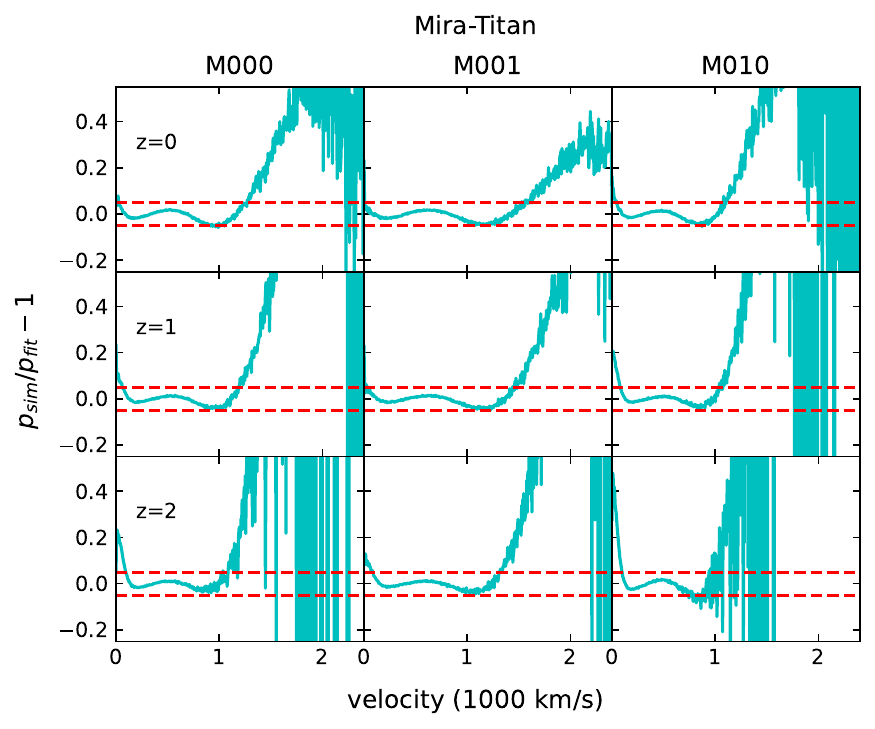}}
   \caption{Left panel: Measured $p(v)$ of halos from models M000, M001, and M010 (Mira-Titan simulations) along with the best-fit Tsallis models (fitted $\kappa_0$ and $\sigma_{\kappa_0}$ in parentheses). Right panel: Relative differences.}
   \label{fig:mtpv}
\end{figure*}

Fitting results are demonstrated in Fig.~\ref{fig:mtpv}. It is apparent that the Tsallis model can fit simulations with different cosmological parameters well. However, the fitting accuracy exhibits slight variations across the models considered. We argue that under gravitational evolution, the velocity distribution of dark matter halos in the simulations gradually evolves from an initial Gaussian distribution toward a non‑Gaussian one. The stronger the gravitational evolution, the stronger the non‑Gaussianity of the system, and consequently, the closer the halo velocity distribution approaches the Tsallis distribution. Therefore, at a given redshift, the goodness of fit of the Tsallis model to the simulation data is in fact related to the strength of the non‑Gaussianity of the system, which is itself modulated by the cosmological parameters.

\begin{figure}[h!]
\centering
   \resizebox{\hsize}{!}{ \includegraphics{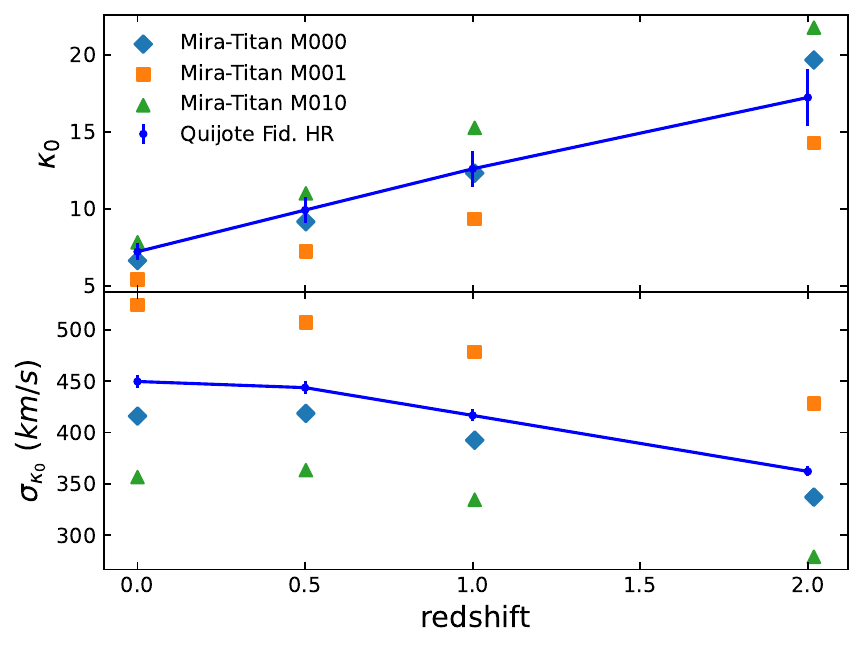}}
   \caption{Redshift dependence of $(\kappa_0,\, \sigma_{\kappa_0})$ in different simulations (results for $z=0.5$, included).}
   \label{fig:zpars}
\end{figure}

We plot the best-fit parameters as functions of redshift in Fig.~\ref{fig:zpars}. It can be seen that, for different cosmological parameters, $\sigma_{\kappa_0}$ differs substantially, while the differences in $\kappa_0$ appear small. Moreover, as redshift approaches zero, the differences in $\kappa_0$ become increasingly smaller, but the differences in $\sigma_{\kappa_0}$ remain almost unchanged. This indicates that the velocity distribution of dark matter halos has strong potential to serve as a diagnostic tool for cosmological models, primarily through its velocity dispersion \citep{Watkins1997, MastersEtal2006}. In contrast, the non‑extensive index, $\kappa_0$, which quantifies the system's departure from thermodynamical equilibrium, exhibits weak discriminating power for cosmological models. Since the main purpose of this paper is to inspect the effectiveness of the Tsallis model, we leave the systematic calibration of the halo velocity distribution for cosmological constraints to future studies.

\section{In the perspective of superstatistics}

\subsection{Theoretical analysis}
The theory of superstatistics states that superposing Gaussians with fluctuating variances yields a family of nonequilibrium distributions, including the Tsallis distribution \citep{Beck2001, BeckCohen2003}. According to the halo model (Eq.~\ref{eq:halomodel}), the peculiar velocities of dark matter halos of a given mass and environment are zero‑mean Gaussian variables, with the mass and environment dependence entering only through the variance \citep{Sheth-Diaferio2001, HamanaEtal2005}. Equation~\ref{eq:halomodel} can therefore be compactly written as
\begin{equation}
p(\vect{v})= \int_0^{+\infty} p_G(\vect{v}; \sigma_G)p(\sigma_G) \rmd \sigma_G = \int_0^{+\infty} p_G(\vect{v}; s)f(s) \rmd s  \ ,
\label{eq:super}
\end{equation}
where $p_G$ is a zero mean Gaussian with a variance $\sigma_G^2=1/s$ and $s$ has a distribution $f(s)$. 

For a $d$-dimensional velocity, $p_G(\vect{v}; s) = (s/2\pi)^{d/2} e^{-v^2s/2}$. If $s$ satisfies the gamma distribution, 
\begin{equation}
f_\Gamma(s;\alpha,\beta)=\frac{\beta^{\alpha+1}}{\Gamma(\alpha+1)} s^\alpha e^{-s\beta}\ ,
\end{equation}
it is straightforward to derive that $p(\vect{v})$ follows a Tsallis distribution with
\begin{equation}
\begin{aligned}
\alpha & = \frac{1}{q-1} - (1+\frac{d}{2})=\kappa_0, \\
\beta  & = \frac{\sigma_q^2}{2(q-1)}=\frac{1}{2} \sigma_{\kappa_0}^2 \kappa_0\ .
\end{aligned}
\label{eq:fgpars}
\end{equation}
An obvious advantage of the formulation is that the dimensional adjustment required in the marginalization procedure for Tsallis statistics becomes automatic. 

If $f(s)$ is not gamma‑distributed, $p(\vect{v})$ belongs to a different class of superstatistics, though it remains analogous to Tsallis statistics. Defining the specific kinetic energy $\epsilon= v^2/2$ and a new distribution function $g(s)=\mathcal{A}^{-1} s^{d/2}f(s)$ with $\mathcal{A}=\int_0^{+\infty} s^{d/2} f(s)\rmd s$, we obtain
\begin{equation}
p(\epsilon)  = \frac{\epsilon^{d/2-1}}{\Gamma(d/2)} \int_0^{+\infty}  s^{d/2} e^{-s\epsilon} f(s) \rmd s = \mathcal{A} \frac{\epsilon^{d/2-1}}{\Gamma(d/2)}   \avg{e^{-s\epsilon}}_g \ .
\label{eq:typeB}
\end{equation}
where $\avg{e^{-s\epsilon}}_g = \int_0^{\infty} e^{-s\epsilon} g(s)\rmd s$, i.e., the average is taken with respect to $g(s)$ rather than $f(s)$ \citep[type-B superstatistics in][]{BeckCohen2003}. For small $\epsilon$, $p(\epsilon)$ matches the Tsallis statistics to order $\avg{\epsilon^2}_g$; for large $\epsilon$, it is mainly controlled by $f(s)$ at small $s$ \citep{TouchetteBeck2005}.

In previous sections, we demonstrated that Tsallis statistics can accurately describe the distribution of halo peculiar velocities for $v\lesssim 1000\ \kms$, with only very mild deviations in the high‑speed regime. It suggests that the corresponding $f(s)$ should be close to a gamma distribution and that the differences between them should not be very large when $s$ is small. Based on this inference, $f(s)$ can be expanded in terms of $f_\Gamma$ and the generalized Laguerre polynomials $L_n^{(\alpha)}(x)=\frac{1}{n!} x^{-\alpha}e^x \frac{\rmd^n}{{\rmd x}^n}(x^{n+\alpha}e^{-x})$:
\begin{equation}
f(s)= f_\Gamma(s; \alpha, \beta) \sum_{n=0}^{+\infty} c_n L_n^{(\alpha)}(\beta s)\ ,
\label{eq:gexp}
\end{equation}
where
\begin{equation}
\begin{aligned}
c_n &= \frac{n! \Gamma(\alpha+1)}{\Gamma(n+\alpha+1)} \int_0^{+\infty} f(s) L_n^{(\alpha)}(\beta s) \rmd s \\
&= \sum_{m=0}^n \frac{n!}{(n-m)! m!} \frac{\Gamma(\alpha+1)}{\Gamma(m+\alpha+1)} (-\beta)^m \avg{s^m}\ ,
\end{aligned}
\label{eq:cn}
\end{equation}
with $\avg{s^m}=\int_0^{+\infty} s^m f(s)\rmd s$. Clearly, $c_0=1$. The expansion is a type of generalized Gram-Charlier expansion involving the gamma distribution. It has been shown to provide good fits to the PDFs of cosmic density fields for dark matter and galaxies \citep{GFE2000, BelEtal2016}.

Substituting the expansion into Eq.~\ref{eq:typeB} gives
\begin{equation}
\begin{aligned}
p(\epsilon) = \sum_{n=0}^{+\infty} & p_n^{(\alpha,\,\beta)}(\epsilon)= \frac{\epsilon^{d/2-1}}{\Gamma(d/2)}  \beta^{-d/2}   \sum_{n=0}^{+\infty} c_n   \frac{\Gamma(n+\alpha+1) }{\Gamma(\alpha+1)}, \\ 
& \sum_{m=0}^n  \frac{ \Gamma(m+\alpha+1+\frac{d}{2}) }{\Gamma(m+\alpha+1)}\frac{(-1)^m}{(n-m)! m!} 
\left(1+ \frac{\epsilon}{\beta}\right)^{-(m+\alpha+1+\frac{d}{2})}\ .
\end{aligned}
\label{eq:ssmodel}
\end{equation}
Owing to the orthogonality of $L_n^{(\alpha)}(x)$,
\begin{equation}
\int_0^{+\infty}p_n^{(\alpha,\, \beta)}(\epsilon)\rmd \epsilon =\begin{cases} 
1\ \quad \text{if } n=0 \ ,\\
0\ \quad \text{otherwise}\ .
\end{cases}
\end{equation} 
Hence, truncating the series at any $n$ automatically yields a normalized approximation. 

Equation~\ref{eq:ssmodel} offers a convenient way to approximate  $p(\vect{v})$ for halos. Truncating the series at order $n^*$ gives
\begin{equation}
\begin{aligned}
&p(\vect{v}) \approx \sum_{n=0}^{n^*} p_n^{(\alpha, \, \beta)}(\vect{v}), \\
&p_n^{(\alpha,\, \beta)}(\vect{v}) =c_n \frac{\Gamma(n+\kappa_0+1)}{\Gamma(\kappa_0+1)}\sum_{m=0}^n
\frac{(-1)^m}{(n-m)!m!} p_{\kappa_m}(\vect{v})\ ,
\end{aligned}
\label{eq:vss}
\end{equation} 
where $p_{\kappa_m}(\vect{v})$ follows Eq.~\ref{eq:kappa} with
\begin{equation}
\begin{aligned}
\kappa_0 &\to \kappa_m=\kappa_0+m =\alpha+m ,\\
\sigma_{\kappa_0}^2 & \to \sigma_{\kappa_m}^2=\sigma_{\kappa_0}^2\frac{\kappa_0}{\kappa_0+m}
=\frac{2\beta}{\alpha+m}\ .
\end{aligned}
\end{equation}
Clearly, $p(\vect{v})$ can be expressed as a sum of the Tsallis distribution functions, and Eq.~\ref{eq:kappa} is just the zero-order approximation. Pragmatically, it is often more convenient to rewrite Eq.~\ref{eq:ssmodel} in orders of $m$ and make a truncation at $m^*$, such that
\begin{equation}
p(\vect{v})\approx\sum_{m=0}^{m^*} \tilde{c}_m p_{\kappa_m}(\vect{v}),
\label{eq:mexp}
\end{equation} 
where the free parameters $\tilde{c}_m$ are fit under the constraint $\sum_{m=0}^{m^*} \tilde{c}_m=1$.

In principle, adding higher-order terms could improve fitting accuracy, but the convergence rate depends on $\alpha$, $\beta$, and moments of $s$ (Eq.~\ref{eq:cn}). If $f(s)$ were known in advance, a good choice of $\alpha$ and $\beta$ would greatly simplify the fitting procedure \citep{MD2010}. However, often one has to derive $f(s)$ from the observed $p(\vect{v})$ by solving Eq.~\ref{eq:super}, which is a Fredholm integral equation of the first kind. Developing a robust and accurate method for the problem is not a simple task. Equations~\ref{eq:vss} and~\ref{eq:mexp} thus serve as applicable models to solve the equation parametrically.

\subsection{Route to refine halo model}

The numerical results presented in the previous section suggest that $f(s)\approx f_{\Gamma}$. Following the halo model approach, it is feasible to directly measure $f(s)$ from N-body simulation data. The local dark matter density contrast, $\tilde{\delta}$, measured within a region centered on each dark halo, can serve as an environmental parameter and is readily obtainable from the simulation. It has been demonstrated that the velocity distribution of halos within the same environment is well approximated by a Gaussian \citep{Sheth-Diaferio2001}; therefore, we only need to compute the velocity variance, $\sigma_G^2$, whose reciprocal defines $s$. With the aid of large N-body simulation suites, $f(s)$, $p(\tilde{\delta})$, the conditional halo mass function $n(M_h|\tilde{\delta})$, and the functional relation between $s$ and $\tilde{\delta}$ can be jointly estimated. These measurements can be used not only to assess the superstatistics interpretation but also to comprehensively calibrate the halo model.

Perhaps the most critical aspect of the halo model in Eq.~\ref{eq:halomodel} is the choice of the characteristic scale, $R$, of the smoothing window used to define $\tilde{\delta}$.  In previous work, this choice has been relatively arbitrary and lacking in theoretical justification. For instance, \citet{Sheth-Diaferio2001} simply illustrated cases with a fixed Gaussian smoothing radius, while \citet{HamanaEtal2005} proposed setting $R$ to be the radius of a top‑hat filter such that $\avg{\tilde{\delta}^2}^{1/2}=0.5(1+z)^{-1/2}$, without a clear physical basis. Consequently, to refine the dark halo model further, we must first identify the most appropriate smoothing scale—including the criteria for determining optimality—and ensure that the adopted scale is applicable to different cosmological models and at different redshifts. This task, however, requires a systematic study using numerical simulations across a range of cosmological models, which goes well beyond the scope of this paper. We therefore leave this investigation for future work.

\section{Summary}
From a series of halo samples drawn from $N$‑body simulations, we find that the Tsallis distribution fits the dark matter halo peculiar velocity distributions well for $\abs{\vect{v}} \lesssim 10^3\ \kms$ at $z\leq 2$, achieving a precision better than $5\%$. Model‑simulation discrepancies are mainly confined to the high‑speed tails and increase with redshift. 

In the $\kappa$ notation, the Tsallis distribution is governed by two parameters: the invariant index $\kappa_0$ and the characteristic velocity $\sigma_{\kappa_0}=\sqrt{2\avg{v^2}/d}$. A smaller $\kappa_0$ indicates stronger non‑Gaussianity and a larger departure from equilibrium. Our analysis of halo samples shows that $\kappa_0$ decreases as $z\to 0$. Mass dependence, while weak, is discernible: low‑mass halos generally exhibit smaller $\kappa_0$ and larger $\sigma_{\kappa_0}$ than their massive counterparts. A key property of the Tsallis distribution is that, while the linear correlation coefficients among the three Cartesian velocity components vanish, the components are not statistically independent. The model further predicts nontrivial quadratic correlations that agree excellently with the behavior of halos in cosmological $N$-body simulations. 

Results from N-body simulations across different cosmologies confirm that the Tsallis model performs robustly against variations in cosmological parameters. Meanwhile, the fitted parameters could be promising candidates for cosmological parameter constraints. While modern observations of the kinematic Sunyaev--Zel'dovich effect and redshift-space distortions predominantly rely on pairwise velocity statistics, the one-point velocity PDF serves as a fundamental prior and building block for these measurements, the non-Gaussianity in the one-point distribution can directly propagate into the pairwise statistics. Therefore, an accurate one-point model is not an alternative to pairwise statistics but rather a necessary theoretical input for them. Based on the numerical results presented here, we conclude that the Tsallis distribution provides an effective and instructive description of the peculiar velocity distribution of halos, particularly at low redshifts, and offers a ready-to-use non-Gaussian prior for halo peculiar velocities, which can be incorporated into kinematic Sunyaev–Zel'dovich analyses and redshift-space power-spectrum modeling.

Motivated by the halo model, we attempted to understand this effectiveness within the framework of superstatistics. In this formalism, the peculiar velocity of a halo is intrinsically Gaussian distributed with a zero mean and variance $\sigma_G^2$. However, the reciprocal of the variance $s=1/\sigma_G^2$ is not fixed. Rather, it follows a distribution that may reflect variations in mass and environment. By expanding the PDF of $s$ in terms of a gamma distribution and generalized Laguerre polynomials, the halo peculiar velocity distribution can indeed be expressed as a sum of a series of Tsallis distribution functions. Since in simulations halo peculiar velocities are well described by a single Tsallis distribution, $f(s)$ must therefore be close to a gamma distribution, making the zeroth‑order approximation already sufficiently accurate. 

In future work, besides refining the halo model and exploring the constraining power of  $(\kappa_0, \sigma_{\kappa_0})$ on cosmological parameters, it would be interesting and worthwhile to investigate the velocity distributions of dark matter and galaxies. Among the dark‑matter–halo–galaxy triad, halos are probably the simplest in terms of peculiar motion. Dark matter and galaxies not only comove with their host halos but also roam within them, which complicates the situation considerably. It is highly plausible that a single Tsallis distribution will prove insufficient.

\begin{acknowledgements}
      This work is supported by the NSFC grant No. 12273049. Ming Li also acknowledges support from the National SKA Program of China (No. 2022SKA0110201) and the National Key Research and Development Program of China (No. 2022YFA1602903). The authors appreciate the anonymous referee’s insightful comments and valuable suggestions, and are grateful to Longlong Feng, Weipeng Lin, Jie Wang, Donghai Zhao, and Pengjie Zhang for fruitful discussions. 
\end{acknowledgements}

\bibliographystyle{aa} 
\bibliography{phv.bib} 
\end{document}